\documentclass[twocolumn,showpacs,preprintnumbers]{revtex4}
\usepackage{amssymb}

\usepackage[dvips]{graphicx}

\begin{document}

\title{Pairing of bosons in the condensed state of the boson-fermion model}
\author{A.S.~Alexandrov}
\affiliation{Department of Physics, Loughborough University,
Loughborough LE11 3TU, United Kingdom}

\begin{abstract}

A two component model of  negative $U$ centers coupled with the
Fermi sea of itinerant fermions is discussed in connection with
high-temperature superconductivity of cuprates, and superfluidity
of atomic fermions. We examine the phase transition and the
condensed state of this boson-fermion model (BFM) beyond the
ordinary mean-field approximation in two and three dimensions. No
pairing of fermions and no condensation are found in
two-dimensions for any symmetry of the order parameter. The
expansion in the strength of the order parameter near the
transition  yields no linear homogeneous term in the
Ginzburg-Landau-Gor'kov equation and a zero upper critical field
in \emph{any}-dimensional BFM, which indicates that previous
mean-field discussions of the model are flawed. Normal and
anomalous
  Green's functions are obtained diagrammatically and analytically in the condensed state of
   a simplest version of 3D BFM. A pairing of
 bosons analogous to the Cooper pairing of
 fermions is found.
There are three coupled condensates in the model, described by the
 off-diagonal  single-particle boson,  pair-fermion and pair-boson
 fields.  These results negate the common wisdom that the boson-fermion model is adequately described by
the BCS theory  at weak coupling.

\end{abstract}

 \maketitle

\bigskip
\section{Introduction}
The experimental \cite {ZHAO,mih,ita,TIM,ega,LANZ} and theoretical
\cite{phil,MOTT,all,gor} evidence for an exceptionally strong
electron-phonon interaction in  novel superconductors is now so
overwhelming that even some advocates of the non-phononic
mechanisms \cite{kiv} admit the fact. A few authors
 (see, for example \cite{SANF,emi,dev,tru}) explored a view that
 the extension of the BCS theory towards
 the strong interaction between electrons and ion vibrations describes
 the phenomenon. In this regime,
pairing takes place in real space due to a polaron collapse of the
Fermi energy \cite{aleran}, or due to a low density of carriers.
At first sight, bipolarons have a mass too large to be mobile.
Indeed, Anderson \cite{pand} introduced small bipolarons as
entirely localised objects explaining some unusual properties of
chalcogenide glasses. However, it has been shown  more recently
that the inclusion of the on-site Coulomb repulsion leads to the
favoured binding of intersite carriers \cite{ALEXAND,CATL}. The
intersite bipolarons can tunnel with an effective mass of about 10
electron masses \cite{ALEXAND,CATL,alekor,tru}, and account for a
high critical temperature  \cite{ale0}.

Soon after Anderson \cite{pand} and Street and Mott \cite{mot}
introduced localized pairs in amorphous semiconductors, a two
component model of negative $U$ centers coupled with the Fermi sea
of itinerant fermions was employed to study superconductivity in
disordered metal-semiconductor alloys \cite{sim,tin}. When the
attractive potential $U$ is large, the model is reduced to
localized hard-core bosons spontaneously decaying into itinerant
electrons and vice versa, different from a non-converting mixture
of mobile charged bosons and fermions \cite{ale3,and}. This
boson-fermion model (BFM) was applied more generally to describe
pairing electron processes with localization-delocalization
\cite{ion}, and  a linear resistivity in the normal state of
cuprates \cite{eli}. The model attracted more attention in
connection with high-temperature superconductors
\cite{lee,ran,ran0,kos,lar,ale,cris,gor,dam,dam2,mic}, as an
alternative to the bipolaron model of cuprates with intrincically
mobile bipolarons \cite{alebook}. In particular, Refs.
\cite{dam,dam2} claimed that 2D BFM with immobile hard-core bosons
is capable to reproduce some physical properties and the phase
diagram of cuprates. BFM has been also adopted for a description
of  superfluidity of atomic fermions scattered into bound
(molecular) states \cite{chio}.

Most studies of BFM  below its transition into a low-temperature
condensed phase applied a
 mean-field approximation (MFA), replacing  zero-momentum boson operators by c-numbers
 and neglecting the boson self-energy  in the density sum rule \cite{lee,ran,kos,lar,dam,dam2,mic,chio}.
  When the bare boson energy is well above the chemical potential, the BCS ground state
  was
  found  with bosons being only virtually excited \cite{lee,ran}.
  MFA led to a conclusion
 that BFM exhibits features compatible with BCS characteristics
 \cite{kos}, and describes a crossover from the BCS-like to local pair  behaviour
 \cite{mic}.  The transition was found  more mean-field-like than
 the usual Bose condensation, i.e. characterized by a relatively
 small value of the fluctuation parameter $Gi$ \cite{lar}.

 At the same time
 our previous study of BFM \cite{ale}  beyond MFA
revealed a crucial effect of the boson self-energy on the normal
state boson spectral function and the transition temperature
$T_{c}$. Ref.\cite{ale} proved that the Cooper pairing of fermions
via virtual bosonic states  is impossible in any-dimensional BFM.
It occurs only simultaneously with the Bose-Einstein condensation
of real bosons. The origin of this simultaneous condensation lies
in a  softening of the boson mode at $T=T_c$ caused by its
hybridization with fermions. The energy of zero-momentum bosons is
renormalized down to \emph{zero}  at $T=T_c$, no matter how weak
the boson-fermion coupling and how large the bare boson energy are
 \cite{ale}.  One can also expect that the boson self-energy should qualitatively
modify the phase transition and the whole
 condensed phase of BFM below $T_c$.

In this paper  the phase transition and the condensed state of
 BFM are examined beyond the ordinary mean-field
approximation in two and three dimensions. It is shown that
$T_{c}=0$ K in the two-dimensional model,  even in the absence of
any Coulomb repulsion,  and the  phase
 transition is never a BCS-like second-order phase transition even in 3D BFM
 because of the complete boson
softening. A closed set of
  equations  for   fermion  and
boson Green's functions (GFs) is derived taking into account the
self-energy effects in the   condensed state of 3D BFM.  There
exist a boson \emph{pair} condensate along with the fermion Cooper
pair and the single-particle boson  condensate in the model.
Remarkably, the Gor'kov expansion \cite{gor2} of GFs in the
strength of the order parameter yields a zero  linear term at
\emph{any} temperature below $T_c$, and zero
 upper critical field.

\section{No Cooper pairing and  condensation in 2D BFM}

2D BFM is defined by the Hamiltonian,
\begin{eqnarray}
H &=&\sum_{{\bf k},\sigma =\uparrow ,\downarrow }\xi _{{\bf k}}c_{{\bf k}%
,\sigma }^{\dagger }c_{{\bf k},\sigma }+E_{0}\sum_{{\bf q}}b_{{\bf q}%
}^{\dagger }b_{{\bf q}}+ \\
&&{\rm g}N^{-1/2}\sum_{{\bf q,k}}\left( \phi _{{\bf k}}b_{{\bf
q}}^{\dagger }c_{-{\bf k}+{\bf q}/2,\uparrow }c_{{\bf k}+{\bf
q}/2,\downarrow }+H.c.\right) ,  \nonumber
\end{eqnarray}
where $\xi _{{\bf k}}=-2t(\cos k_{x}+\cos k_{y})-\mu $ is the 2D
energy spectrum of fermions, $E_{0}\equiv \Delta _{B}-2\mu $ is
the bare boson energy with respect to their chemical potential
$2\mu $, ${\rm g}$ is the
magnitude of the anisotropic hybridization interaction, $\phi _{{\bf k}%
}=\phi _{-{\bf k}}$ is the anisotropy factor, and $N$ is the
number of cells. Here and further I take $\hbar=c=k_B=1$ and the
lattice constant $a=1$. Ref. \cite{dam} argued that
'superconductivity is induced in this
model from the anisotropic charge-exchange interaction (${\rm g}\phi _{{\bf k%
}}$) between the conduction-band fermions and the immobile
hard-core bosons', and 'the on-site Coulomb repulsion  competes
with this pairing' reducing the critical temperature $T_{c}$ less
than by 25\%. Also it has been argued \cite{dam2}, that the
calculated upper critical field of the model fits well the
experimental results in cuprates.

 Here I show that
$T_{c}=0$ K in the two-dimensional model, Eq.(1), even in the
absence of any Coulomb repulsion,  and the mean-field
approximation is meaningless for any-dimensional BFM because of
the complete boson softening.

Replacing boson operators by $c$-numbers for ${\bf q}=0$ in Eq.(1)
one obtains  a linearised BCS-like equation for the fermion
order-parameter (the gap function) $\Delta _{{\bf k}}$,
\begin{equation}
\Delta _{{\bf k}}=\frac{{\rm \tilde{g}}^{2}\phi _{{\bf k}}}{E_{0}N}\sum_{%
{\bf k}^{\prime }}\phi _{{\bf k}^{\prime }}{\frac{\Delta _{{\bf k}^{\prime
}}\tanh (\xi _{{\bf k}^{\prime }}/2T)}{{2\xi _{{\bf k}^{\prime }}}},%
}
\end{equation}
with the coupling constant  ${\rm \tilde{g}}^{2}={\rm
g}^{2}(1-2n^{B})$, renormalized by the hard-core effects. Using a
two-particle fermion vertex part in the Cooper channel one can
prove that this equation is perfectly correct even beyond the
conventional non-crossing approximation \cite{ale}. The problem
with MFA does not stem  from this BSC-like equation, but from an
incorrect definition of the bare boson energy with respect to the
chemical potential, $E_{0}(T)$. This energy is determined by the
atomic density of bosons ($n^{B}$)  as (Eq.(9) in Ref. \cite{dam})
\begin{equation}
\tanh \frac{E_{0}}{2T}=1-2n^{B}.
\end{equation}
While Eq.(2) is correct, Eq.(3) is incorrect because the boson
self-energy $\Sigma _{b}({\bf q},\Omega)$ due to the same
hybridization  interaction is missing. At first sight \cite{dam}
the self-energy  is small in comparison to the kinetic energy of
fermions, if ${\rm g}$ is small. However $\Sigma _{b}(0,0)$
diverges logarithmically at zero temperature \cite{ale}, no matter
how week the interaction is. Therefore it should be kept in the
density sum-rule, Eq.(3). Introducing the boson Green's function
\begin{equation}
D({\bf q},\Omega)=\frac{1-2n^{B}}{i\Omega -E_{0}-\Sigma _{b}({\bf q}%
,\Omega)}
\end{equation}
one must replace incorrect Eq.(3) by

\bigskip
\begin{equation}
-{\frac{T}{{N}}}\sum_{{\bf q},n}e^{i\Omega \tau }D({\bf q}%
,\Omega)=n^{B},
\end{equation}
where $\tau =+0$, and $\Omega =2\pi Tn$ ($n=0,\pm 1,\pm 2...$).

\begin{figure}
\begin{center}
\includegraphics[angle=-90,width=0.67\textwidth]{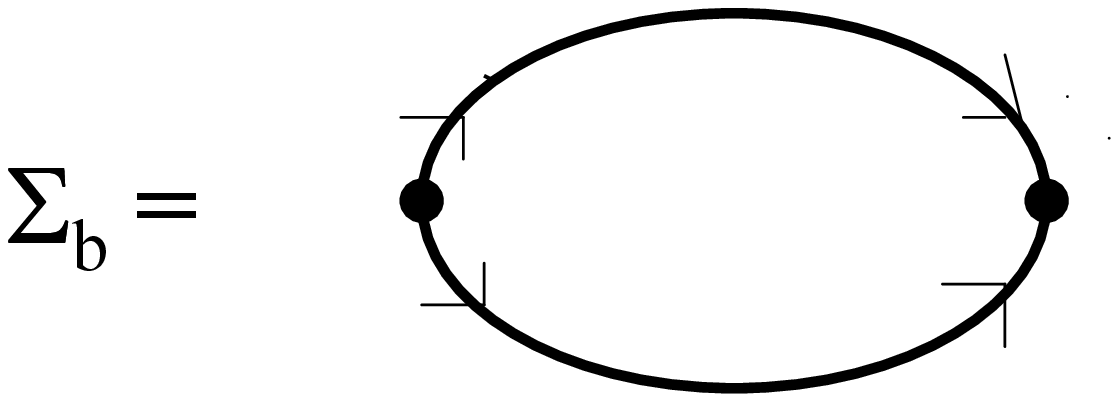}
 \caption{Diagram for the  boson self-energy. Solid lines are the
 fermion GFs. Vertex (dot) corresponds to the hybridization
 interaction.}
\end{center}
\end{figure}

The divergent (cooperon) contribution to $\Sigma _{b}({\bf
q},\Omega)$ is given by Fig.1 \cite{ale},
\begin{eqnarray}
&&\Sigma _{b}({\bf q},\Omega)=-\frac{{\rm \tilde{g}}^{2}}{2N}\sum_{{\bf %
k}}\phi _{{\bf k}}^{2}\times  \\
&&\frac{\tanh [\xi _{{\bf k-q}/2}/(2T)]+\tanh [\xi _{{\bf k+q}/2%
}/(2T)]}{\xi _{{\bf k-q}/2}+\xi _{{\bf k+q}/2}-i\Omega}, \nonumber
\end{eqnarray}
so that one obtains
\begin{equation}
\Sigma _{b}({\bf q},0)=\Sigma _{b}(0,0)+\frac{q^{2}}{2M^{\ast }}+{\cal O}%
(q^{4})
\end{equation}
for small ${\bf q}$ and any anisotropy factor compatible with the
point-group symmetry of the cuprates. Here $M^{\ast }$ is the
boson mass, calculated analytically in Ref.\cite{ale} for the
isotropic exchange
interaction and parabolic fermion band dispersion (see also Ref.\cite{cris}%
). The BCS-like equation (2) has a nontrivial solution for $\Delta
_{{\bf k}} $ at $T=T_c$, if
\begin{equation}
E_{0}=-\Sigma _{b}(0,0).
\end{equation}
Substituting Eq.(7) and Eq.(8) into the sum-rule, Eq.(5), one
obtains a logarithmically divergent integral with respect to ${\bf
q}$, and
\begin{equation}
T_{c}=\frac{const}{\int_{0}dq/q}=0.
\end{equation}
The devastating result, Eq.(9) is a direct consequence of the
well-known theorem, which states that BEC is impossible in 2D.

One may erroneously believe that MFA results\cite{dam,dam2}  can
be still applied in three-dimensions, where BEC is possible.
However, increasing dimensionality does not make MFA a meaningful
approximation for the boson-fermion model. This approximation
leads to a naive conclusion that a BCS-like superconducting state
 occurs below the
 critical temperature   $T_{c}\simeq \mu \exp\left( -{%
E_{0}/z_c}\right) $ via fermion pairs being \emph{virtually}
excited into
 $unoccupied$ bosonic states \cite{lee,ran}.  Here $z_c=\tilde{g}^{2}N(0)$ and
$N(0)$ is the density of states (DOS) in the fermionic band near
the Fermi level $\mu $. However,  the Cooper pairing of fermions
 is impossible via virtual unoccupied bosonic states also in 3D BFM. Indeed,
Eqs.(2,8) do not depend on the dimensionality, so that the
analytical continuation of Eq.(4) to real frequencies $\omega$
yields the partial boson DOS as $\rho(\omega)=(1-2n_B)
\delta(\omega)$ at $T=T_c$ and ${\bf q}=0$ in any-dimensional BFM.
The Cooper pairing may occur only simultaneously with the
Bose-Einstein condensation of real bosons in the exact theory of
3D BFM \cite{ale}. The origin of the simultaneous condensation of
the fermionic and bosonic fields in 3D BFM lies in the  softening
of the boson mode at $T=T_c$ caused by its hybridization with
fermions.

Taking into account the boson damping and dispersion shows that
the boson spectrum significantly changes for all momenta.
Continuing the self-energy, Eq.(6) to real frequencies yields  the
damping (i.e. the imaginary part of the self-energy) as \cite{ale}
\begin{equation}
\gamma({\bf q},\omega)={\pi z_c\over{4q\xi}} \ln
\left[{\cosh(q\xi+\omega/(4T_{c}))\over{\cosh(-q\xi+\omega/(4T_{c}))}}\right],
\end{equation}
where $\xi=v_F/(4T_{c})$ is a coherence length, and $v_F$ is the
Fermi velocity. The damping is significant when $q\xi<<1$. In this
region $\gamma({\bf q},\omega)=\omega\pi z_{c}/(8T_{c})$ is
comparable or even larger than  the boson energy $\omega$. Hence
bosons look like overdamped diffusive modes, rather than
quasiparticles in the long-wave limit \cite{ale,cris}, contrary to
the erroneous conclusion of Ref.\cite{ran0}, that there is 'the
onset of coherent free-particle-like motion of the bosons' in this
limit. Only outside the long-wave   region, the damping becomes
small. Indeed, using Eq.(10) one obtains $\gamma({\bf
q},\omega)=\omega \pi z_{c}/(2qv_F)<< \omega$, so that bosons at
 $q >>1/\xi$ are well defined quasiparticles
 with a logarithmic dispersion, $\omega(q)=z_c \ln(q
\xi)$ \cite{ale}.  Hence the boson energy disperses over the whole
energy interval from zero up to $E_0$.

The main mathematical problem with MFA in 3D also stems from the
density sum rule, Eq.(5) which determines the chemical potential
of the system and consequently the bare boson energy $E_{0}(T)$ as
a function of temperature. In the framework of MFA one takes the
bare boson energy   in Eq.(2) as a temperature independent
parameter, $E_0=\tilde{g}^2N(0)\ln (\mu/T_c)$ \cite{lar}, or
determines it from the conservation of the total number of
particles  neglecting the boson self-energy, Eq.(3)
\cite{ran,dam,mic,chio}. Then Eq.(2) looks like the conventional
linearized Ginzburg-Landau-Gor'kov equation \cite{gor2}  with a
negative coefficient $\alpha \propto T-T_c$ at $T<T_c$ in the
linear term. Then one concludes that the phase transition is
almost the conventional BCS-like transition, at least at $E_0\gg
T_c$ \cite{lee,ran,lar}. These findings are mathematically and
physically erroneous. Indeed, the term of the sum in Eq.(5) with
$\Omega_n=0$ is given by the integral
\begin{equation}
T\int {d{\bf q}\over{2\pi^3}}{1\over{E_0+\Sigma_b({\bf q},0)}}.
\end{equation}
 The integral converges, if and
only if $ E_0\geqslant -\Sigma_b(0,0)$. In fact,
\begin{equation}
E_0+\Sigma_b(0,0)=0
\end{equation}
 is strictly zero in the
Bose-condensed state, because $\mu_b=-[E_0+\Sigma_b(0,0)]$
corresponds to the boson chemical potential relative to the lower
edge of the boson energy spectrum. More generally, $\mu_b=0$
corresponds to the appearance of the Bogoliubov-Goldstone mode due
to a broken symmetry  below $T_c$. This exact result makes the BSC
equation (2) simply an identity \cite{ale} with  $\alpha(T) \equiv
0$ at any temperature below $T_c$. On the other hand, MFA
 violates the density
sum-rule, predicting the wrong negative $\alpha(T)$ below $T_c$.

 Since $\alpha(T)=0$, one may expect that the conventional upper
 critical field, $H_{c2}(T)$ is zero in BFM. To determine
 $H_{c2}(T)$ and explore the condensed phase of 3D BFM, one can
  apply the Gor'kov formalism \cite{gor2}, as described below.

\section{Normal and anomalous Green's functions of 3D BFM: pairing of bosons}

Let us now explore a simplified version of 3D BFM  in an external
magnetic field ${\bf B= \nabla \times A}$ neglecting the hard-core
effects,
\begin{eqnarray}
H&=&\int d{\bf r}\sum_s \psi _{s}^{\dagger }({\bf r})\hat{h}({\bf
r})\psi _{s}({\bf r}) +g[\phi({\bf r})\psi _{\uparrow }^{\dagger
}({\bf r})\psi _{\downarrow }^{\dagger }({\bf
r})+H.c.]\nonumber\\
&+&E_0\phi^{\dagger}({\bf r})\phi({\bf r}),
\end{eqnarray}
where $\psi_{s}({\bf r})$ and $\phi({\bf r})$ are fermionic and
bosonic fields, $s=\uparrow, \downarrow$ is the spin,
$\hat{h}({\bf r)=}-[\nabla +ie{\bf A(r)]}^{2}/(2m)-\mu$ is the
fermion kinetic energy operator. Here  the volume of the system is
taken as $V=1$.

The Matsubara field operators, $Q=\exp (H\tau )Q({\bf r)}\exp
(-H\tau ), \bar{Q}=\exp(H\tau) Q^{\dagger}({\bf r)}\exp (-H\tau )$
($Q\equiv\psi_s, \phi$) evolve with the imaginary time
$-1/T\leqslant \tau \leqslant 1/T$ as

\begin{eqnarray}
-\frac{\partial \psi _{\uparrow }({\bf r},\tau )}{\partial \tau } &=&\hat{h}(%
{\bf r)}\psi _{\uparrow }({\bf r},\tau )+g \phi({\bf r},
\tau)\bar{\psi} _{\downarrow }({\bf r,}\tau ), \\
\frac{\partial \bar{\psi} _{\downarrow }({\bf r,}\tau )}{\partial
\tau } &=&\hat{h}^{\ast }({\bf r)}\bar{\psi} _{\downarrow }({\bf
r,}\tau )-g \bar{\phi}({\bf r},\tau)\psi _{\uparrow }({\bf r,}\tau
), \\ -\frac{\partial \phi ({\bf r,}\tau )}{\partial \tau} &=& E_0
\phi({\bf r,}\tau )+g\psi _{\downarrow }({\bf r,}\tau )\psi
_{\uparrow }({\bf r,}\tau ).
\end{eqnarray}
The  theory of the condensed state can be formulated with the
 normal and anomalous fermion GFs \cite{gor2}, $ {\cal
G}({\bf r,r}^{\prime },\tau)=- \langle T_{\tau
}\psi _{s}({\bf r},\tau )\bar{\psi} _{s}({\bf r}^{\prime }{\bf ,}%
0)\rangle$, ${\cal F}^{+}({\bf r,r}^{\prime },\tau)= \langle
T_{\tau }\bar{\psi} _{\downarrow }({\bf r,}\tau) \bar{\psi}
_{\uparrow }({\bf r}^{\prime
},0)\rangle$, respectively, where the operation $%
T_{\tau }$ performs the time ordering. Fermionic and bosonic
fields condense simalteneously \cite{ale}. Following Bogoliubov
\cite{bog} the bosonic  condensate is described by separating a
large matrix element $\phi_{0} ({\bf r})$ in $\phi({\bf r},\tau) $
as a number, while the remaining part $\tilde{\phi}({\bf r},\tau)$
describes a  supracondensate field, $ \phi ({\bf r},\tau)=\phi
_{0}({\bf r})+\tilde{\phi}({\bf r},\tau)$. Then using Eq.(16) one
obtains
\begin{equation}
g\phi_0({\bf r})= \Delta({\bf r})\equiv-{g^2\over{E_0}}{\cal
F}({\bf r,r},0+),
\end{equation}
where ${\cal F}({\bf r,r}^{\prime },\tau)= \langle T_{\tau }\psi
_{\downarrow }({\bf r,}\tau) \psi _{\uparrow } ({\bf r}^{\prime
},0)\rangle$. The equations for  GFs are obtained by using Eqs.
(14-16) and the  diagrammatic technique \cite{abr} in the
framework of the non-crossing approximation \cite{ref}, as shown
in Fig.2 and Fig.3.

An important novel feature of BFM is a pairing of supracondensate
bosons, caused by their hybridization with the fermionic
condensate, as follows from the last diagram in Fig.3. Hence, one
has to introduce an $anomalous$ supracondensate boson GF, ${\cal
B}^{+}({\bf r,r}^{\prime },\tau)=\langle T_{\tau
}\bar{\tilde{\phi}}({\bf r,}\tau) \bar{\tilde {\phi}}({\bf
r}^{\prime },0)\rangle$ along with the normal boson GF, ${\cal
D}({\bf r,r}^{\prime },\tau)=- \langle T_{\tau
}\tilde{\phi} ({\bf r},\tau )\bar{\tilde{\phi}}({\bf r}^{\prime }{\bf ,}%
0)\rangle$.
\begin{figure}
\begin{center}

\includegraphics[angle=-90,width=0.57\textwidth]{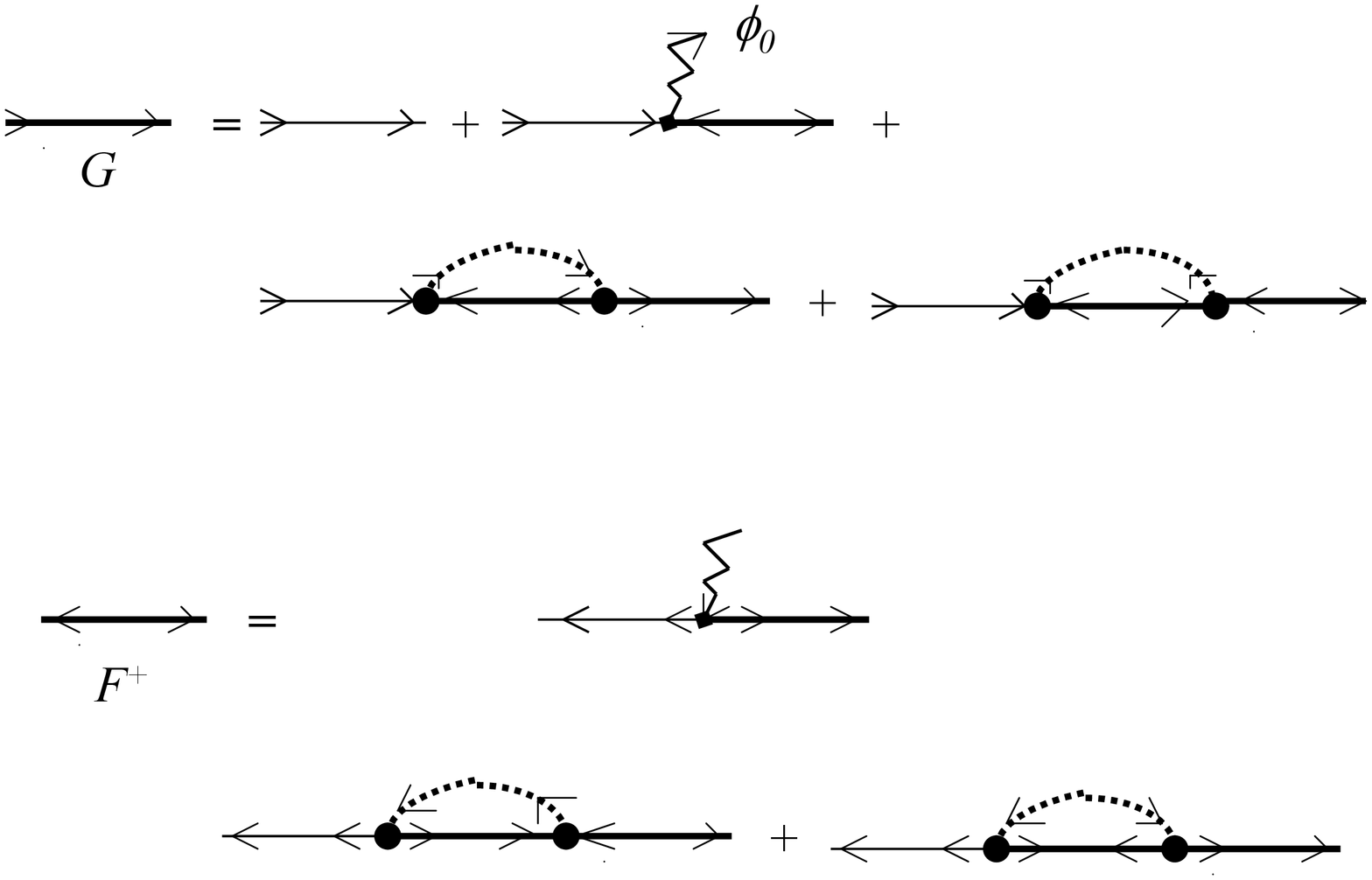}
\caption{Diagrams for the normal and anomalous fermion GFs.
Zig-zag arrows represent the single-particle Bose condensate
$\phi_0$, dotted lines are the boson GFs.}
\end{center}
\end{figure}
The diagrams, Fig.2 and Fig.3, are transformed into analytical
equations for the time Fourier-components of the fermion GFs with
the Matsubara frequencies $\omega=\pi T (2n+1)$ ($n=0, \pm 1,\pm
2,...$) as
\begin{eqnarray}
&&[i\omega -\hat{h}({\bf r)}]{\cal G}_{\omega}({\bf r,r}^{\prime
})
 = \delta ({\bf r-r}^{\prime })-\Delta ({\bf r)}{\cal
F}_{\omega}^{+}({\bf r,r}^{\prime } ) \cr
 &-& g^2 T
\sum_{\omega^{\prime}} \int d{\bf x}{\cal
G}_{-\omega^{\prime}}({\bf x,r}){\cal
D}_{\omega-\omega^{\prime}}({\bf r,x}){\cal G}_{\omega}({\bf
x,r}^{\prime})
 \cr &-& g^2T
 \sum_{\omega^{\prime}} \int d{\bf
x}{\cal F}^{+}_{\omega'}({\bf r,x}){\cal
B}_{\omega+\omega^{\prime}} ({\bf r,x}){\cal F}^{+}_{\omega}({\bf
x,r^{\prime}}),
\end{eqnarray}

\begin{eqnarray}
&&[-i\omega -\hat{h}^{\ast}({\bf r)}]{\cal F}^{+}_{\omega}({\bf
r,r}^{\prime }) = \Delta^{\ast} ({\bf r)}{\cal G}_{\omega}({\bf
r,r}^{\prime } ) \cr
 &-& g^2 T
\sum_{\omega^{\prime}} \int d{\bf x}{\cal
G}_{\omega^{\prime}}({\bf r,x}){\cal
D}_{\omega^{\prime}-\omega}({\bf x,r}){\cal F}^{+}_{\omega}({\bf
x,r}^{\prime}) \cr &+& g^2T
 \sum_{\omega^{\prime}} \int d{\bf
x}{\cal F}_{-\omega'}({\bf r,x}){\cal
B}^{+}_{-\omega-\omega^{\prime}} ({\bf r,x}){\cal G}_{\omega}({\bf
x,r^{\prime}}) \nonumber
\end{eqnarray}
 and,
\begin{eqnarray}
&&(i\Omega -E_0) {\cal D}_{\Omega}({\bf r,r}^{\prime }) = \delta
({\bf r-r}^{\prime })\cr &-& g^2 T \sum_{\omega^{\prime}} \int
d{\bf x}{\cal G}_{\omega^{\prime}}({\bf r,x}){\cal
G}_{\Omega-\omega^{\prime}}({\bf r,x}){\cal D}_{\Omega}({\bf
x,r}^{\prime}) \cr &-& g^2T
 \sum_{\omega^{\prime}} \int d{\bf
x}{\cal F}_{\omega'}({\bf r,x}){\cal F}_{\Omega-\omega^{\prime}}
({\bf r,x}){\cal B}^{+}_{\Omega}({\bf x,r}^{\prime}),
\end{eqnarray}
\begin{eqnarray}
&&(-i\Omega -E_0){\cal B}^{+}_{\Omega}({\bf r,r}^{\prime }) = \cr
&&  g^2T
 \sum_{\omega^{\prime}} \int d{\bf
x}{\cal F}_{-\omega'}^{+}({\bf r,x}){\cal
F}_{-\Omega+\omega^{\prime}}^{+}({\bf r,x}){\cal D}_{\Omega}({\bf
x,r^{\prime}}) \cr &-& g^2 T \sum_{\omega^{\prime}} \int d{\bf
x}{\cal G}_{-\omega^{\prime}}({\bf x,r}){\cal
G}_{\omega^{\prime}-\Omega}({\bf x,r}){\cal B}^{+}_{\Omega}({\bf
x,r}^{\prime}). \nonumber
\end{eqnarray}
for the boson GFs  with  ${\cal B}({\bf r,r}^{\prime
},\tau)=\langle T_{\tau }\tilde{\phi}({\bf r,}\tau) \tilde
{\phi}({\bf r}^{\prime },0)\rangle$.

\begin{figure}
\begin{center}

\includegraphics[angle=-90,width=0.57\textwidth]{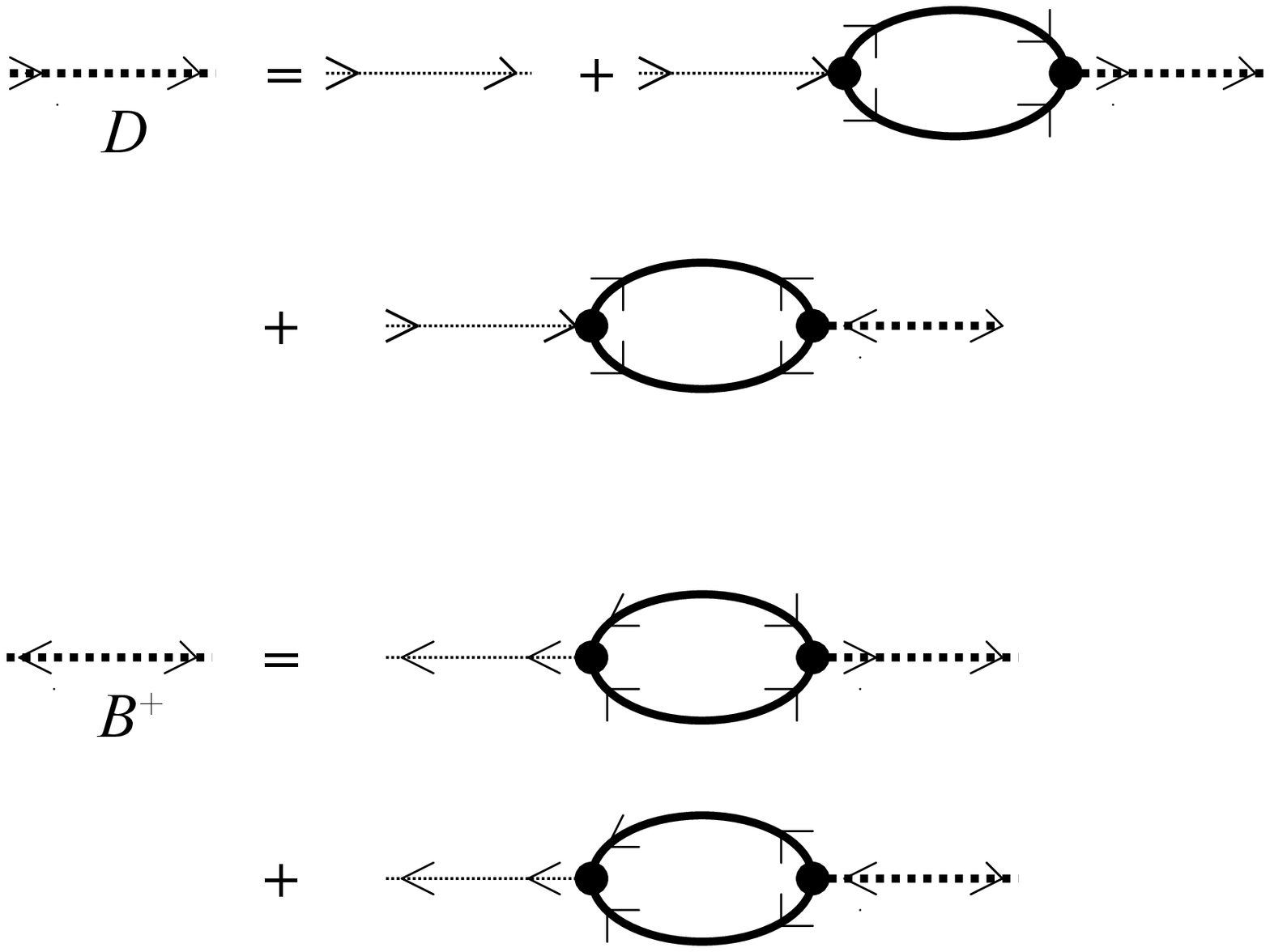}
 \caption{Diagrams for the supracondensate boson
 GFs. The Cooper-pairing of fermions  leads to the Cooper-pair-like boson condensate,
 described by the boson anomalous GF, ${\cal
B}^{+}$.}
\end{center}
\end{figure}

\section{Gor'kov expansion}

These equations can be formally solved in the homogeneous case
without the external field, ${\bf A}=0$. Transforming into the
momentum space yields GFs' time-space Fourier components as
\begin{eqnarray}
{\cal G}({\bf k},\omega) &=-&{\frac{i\tilde{\omega}^{\ast} +\xi
_{\bf k}}{ |i\tilde{\omega} -\xi _{\bf
k}|^2+|\tilde{\Delta}({\bf k}, \omega)|^{2}}},\\
{\cal F}^{+}({\bf k}, \omega) &=&{\frac{\tilde{\Delta}^{\ast}({\bf
k}, \omega)}{ |i\tilde{\omega} -\xi _{\bf
k}|^2+|\tilde{\Delta}({\bf k}, \omega)|^{2}}} ,
\end{eqnarray}
and
\begin{eqnarray}
{\cal D}({\bf q},\omega) &=&-{\frac{i\tilde{\Omega}^{\ast} +E_0}
{|i\tilde{\Omega} -E_0|^2+|\Gamma({\bf q}, \Omega)|^{2}}},
 \\
{\cal B}^{+}({\bf q}, \omega) &=&{\frac{\Gamma^{\ast}({\bf q},
\Omega) }{ |i\tilde{\Omega} -E_0|^2+|\Gamma({\bf q},
\Omega)|^{2}}},
\end{eqnarray}
where $\tilde{\omega}\equiv\omega+i\Sigma_f({\bf k}, \omega)$, $
\tilde{\Omega}\equiv\Omega+i\Sigma_b({\bf q}, \Omega)$, and $\xi
_{\bf k}=k^{2}/(2m)-\mu$.  The fermionic order parameter,
renormalised with respect to the mean-field $\Delta$ due to the
formation of the boson-pair condensate, is given by
\begin{equation}
\tilde{\Delta}({\bf k}, \omega)=\Delta + g^2T
 \sum_{\omega^{\prime}}\int {d {\bf q}\over{2\pi^3}} {\cal F}^{+}({\bf k}-{\bf q},\omega'
){\cal B}({\bf q},\omega+\omega^{\prime}),
\end{equation}
and the boson-pair order parameter, generated by the hybridization
with the fermion Cooper pairs, is
\begin{equation}
 \Gamma({\bf
q}, \Omega)= g^2T \sum_{\omega^{\prime}}\int {d {\bf
k}\over{2\pi^3}} {\cal F}({\bf k},\omega'){\cal F}({\bf q}-{\bf
k},\Omega-\omega^{\prime}).
\end{equation}
  Hence,
there are three coupled condensates in the model described by the
 off-diagonal fields $g\phi_0$, $\tilde{\Delta}$, and
$\Gamma$, rather than two, as in MFA. At low temperatures all of
them have about the same magnitude, as the fermion, Fig.4, and
boson, Fig.1, self-energies,
\begin{equation}
\Sigma_f({\bf k}, \omega)=-g^2T
 \sum_{\omega^{\prime}}\int {d {\bf q}\over{2\pi^3}} {\cal G}({\bf q}-{\bf k},-\omega'){\cal
D}({\bf q},\omega-\omega^{\prime}),
\end{equation}
\begin{equation}
\Sigma_b({\bf q}, \Omega)=-g^2T
 \sum_{\omega^{\prime}}\int {d {\bf q}\over{2\pi^3}} {\cal G}({\bf k},\omega'){\cal
G}({\bf q}-{\bf k},\Omega-\omega^{\prime}),
\end{equation}
 respectively.

\begin{figure}
\begin{center}
\includegraphics[angle=-90,width=0.67\textwidth]{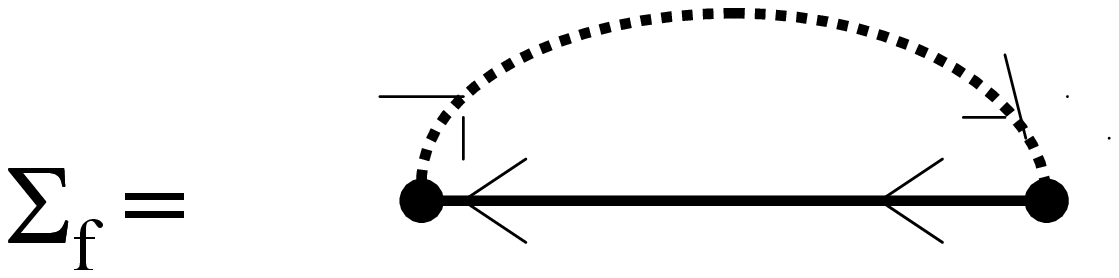}
 \caption{The fermion self-energy}
\end{center}
\end{figure}

On the other hand, when the temperature is   close to $T_c$ (i.e.
$T_c-T \ll
 T_c$),
   the boson pair condensate  is weak compared with the single-particle boson and the Cooper pair condensates.
  In this temperature range
$\Gamma$, Eq.(25) is of the second order in $\Delta$, $\Gamma
\propto \Delta^{2}$, so that the anomalous boson GF can be
neglected, since $\Delta$ is small. The fermion self-energy,
Eq.(26) is a regular function of $\omega$ and ${\bf k}$, so that
it can be absorbed in the renormalized fermion band dispersion.
Then the fermion normal and anomalous GFs, Eqs.(20,21) look like
the familiar GFs  of the BCS theory, and one can apply the Gor'kov
expansion \cite{gor2}  in powers of $\Delta ({\bf r)}$ to describe
the condensed phase of BFM in the magnetic field near the
transition.  Using Eq.(18)  one obtains
 to the terms linear in $\Delta $
\begin{equation}
\Delta^{\ast} ({\bf r)} ={g^2\over{E_0}}T\sum_{\omega _{n}}\int d{\bf x}{\cal G}%
_{-\omega _{n}}^{(n)}({\bf x,r})\Delta ^{\ast }({\bf x)}{\cal
G}_{\omega _{n}}^{(n)}({\bf x,r}).
\end{equation}
The spatial variations of the vector potential  are small near the
transition.  If ${\bf A}({\bf r})$ varies slowly, the normal state
GF, ${\cal G}_{\omega }^{(n)}({\bf r,r}^{\prime })$ differs from
the zero-field normal state GF, ${\cal G}_{\omega
}^{(0)}(\bf{r-r^{\prime}} )$ only by a phase \cite{gor2} ${\cal
G}_{\omega }^{(n)}({\bf r,r}^{\prime })=\exp [-ie{\bf A(r)}\cdot
(\bf{r-r}^{\prime })]{\cal G}_{\omega }^{(0)}(\bf{r-r}^{\prime }
)$. Expanding all quantities  near the point ${\bf x=r}$ in
Eq.(28) up to the second order in ${\bf x-r} $ inclusive, one
obtains the linearised  equation for the fermionic order parameter
as
\begin{equation}
 \gamma [\nabla -2ie{\bf
A(r)]}^{2}\Delta({\bf r})=\alpha \Delta({\bf r}),
\end{equation}
where
\begin{equation}
\alpha = 1+{\Sigma_b(0,0)\over{E_0}}\approx 1-
{g^2N(0)\over{E_0}}\ln {\mu\over{T}},
\end{equation}
and $\gamma \approx 7\zeta (3)v_{F}^{2}g^2N(0)/(48\pi^2
T^{2}E_0)$.

 \section{Conclusion}
 The coefficient $\alpha(T)$ disappears in Eq.(29), since $E_{0}=-\Sigma _{b}(0,0)$ at and \emph{below} $T_c$, Eq.(12).
It means that the  phase
 transition is never a BCS-like second-order phase transition
 even at large $E_0$ and small $g$. In fact, the
 transition  is driven by the Bose-Einstein condensation of \emph{
 real} bosons with ${\bf q}=0$, which occur  due to the complete
 softening of their spectrum at  $T_c$.
 Remarkably, the conventional upper critical field, determined as the field, where a non-trivial
 solution of the \emph{linearised} Gor'kov equation (29)  occurs, is
 zero in BFM, $H_{c2}(T)=0$. It is not  a finite $H_{c2}(T)$
 found in Ref. \cite{dam} using MFA.

 This qualitative failure of MFA  might be rather unexpected, if
one believes that bosons in Eq.(1) play the same role as phonons
in the BCS superconductor. This is not the case for two reasons.
The first one is  the density sum-rule, Eq.(5), for bosons which
is not applied to phonons. The second being that the boson
self-energy is given by the divergent (at $T=0$) Cooperon diagram,
while the self-energy of phonons is finite at small coupling.

In the homogeneous case $\Delta(T)$ should be determined from
Eq.(5)
 rather than from the BCS-like equation (2), which
is actually the identity. To get an insight
 into the magnetic properties of the condensed phase
one has to solve Eqs.(18,19) and Eq.(5) keeping the non-linear
terms. Even
 at  temperatures well below $T_c$ the condensed state is fundamentally
 different from the MFA ground state, because of the pairing of
 bosons. The latter is similar to the Cooper-like pairing of
 supracondensate $^{4}He$ atoms \cite{pas}, proposed  as an explanation of the small density of the single-particle
 Bose condensate in superfluid Helium-4.   The pair-boson condensate
 should significantly
modify the thermodynamic
  properties of the condensed BFM
  compared with the MFA predictions.
   The common wisdom
 that at weak coupling  the boson-fermion model is adequately described by the BCS
 theory, is therefore negated by our theory.

I highly appreciate enlightening   discussions with A.F. Andreev,
L.P. Gor'kov, V.V. Kabanov, A.I. Larkin,   A.P. Levanyuk, R.
Micnas,  and S. Robaszkiewicz,
and support by the Leverhulme Trust
(UK) via Grant F/00261/H.


\begin{thebibliography}{90}
\bibitem{ref} The crossing diagrams are small  as $g/\mu
\ll 1$ \cite{ale}.
\end{thebibliography}
\end{document}